# Fractional Oscillator - Harmonic Oscillator with Memory Effects

*Vishwamittar[a], Yashika Taneja[b] and Nipun Ahuja[b]*

[a] *Department of Physics, Panjab University, Chandigarh – 160014.*
[b] *Department of Mechanical Engineering, Delhi Technological University, Delhi – 110042.*
email addresses: vm@pu,ac.in, taneja.ysh@gmail.com, and ahuja.nipun@gmail.com.

**Abstract**

The importance of fractional time-derivative to take care of memory effects has been brought out by considering the example of a simple oscillator.

## 1. Introduction

The calculus, to which students are exposed to at school level, deals with derivatives of integer order, expressed as $\frac{d^n y(x)}{dx^n}$. Natural question that can come to one's mind is: Will the idea of derivative be valid if it is extended to non-integer value of $n$? In fact, such a query formed part of correspondence between L'Hospital, one of the main developers of calculus, and Leibniz (one of the inventors of this branch of mathematics), nearly 325 years back in 1695. In one letter, the former sought latter's opinion about the meaning of $\frac{d^n y(x)}{dx^n}$ if $n$ had fractional value 1/2. It was on Sept 30, 1695, that Leibniz sent the answer adding that 'this was an apparent paradox', and made a prophetic statement that 'one day, useful consequences would be drawn from this'. Initially, only occasional reference was made to fractional derivatives by some mathematicians and serious efforts to develop the subject in a logical manner began in the first half of the 19$^{th}$ century, and this got contributions from many prominent mathematicians till the middle of the 20$^{th}$ century. Interestingly, many of them introduced their own formulae for fractional (and even any arbitrary real non-integer or complex order) derivative and integral, with strange properties, following different approaches.

A noteworthy development from the point-of-view of future applications of the subject, particularly in physical and engineering problems, occurred in 1967 when Caputo put forward his definition of the fractional time-derivative (FD) of order $\alpha$ as

$$_a D^\alpha f(t) = \frac{1}{\Gamma(n-\alpha)} \left[ \int_a^t (t-t')^{(n-\alpha)-1} \frac{d^n f(t')}{dt'^n} dt' \right]; \qquad n \in \mathbb{N} \text{ and } n-1 < \alpha \leq n. \qquad (1)$$

Here, $\Gamma(z)$ is Euler's gamma function defined as $\Gamma(z) = \int_0^\infty t^{z-1} e^{-t} dt$, where $z$ is a complex number such that its real part is positive. This satisfies the property $\Gamma(z+1) = z\Gamma(z)$, and in particular, $\Gamma(1/2) = \sqrt{\pi}$ and $\Gamma(n+1) = n!$. It is important to note that FD involves integral of $\frac{d^n f(t')}{dt'^n}$ getting contribution with weight $(t-t')^{n-\alpha-1}$ from initial time $a$ to the instant of observation $t$. In other words, the FD at time $t$ is collection of weighted values in the earlier times so that it includes the influence of the past behaviour of the function as if it had some memory to remember the states through which it had passed and, therefore, it is nonlocal in nature. This is in contrast with conventional integer-order derivatives which are evaluated only for the value of t at the time of interest making it local. Thus, fractional order derivatives are useful in describing time-evolution of physical systems with memory effects of all types. It may be pointed out that the value of the order of derivative provides a measure of the memory of the system. However, it must be emphasized that the (physical) system described by such a function does not have any hidden intelligence but only delayed effects of collisions, interactions, causative forces, etc.

The concepts and methods of fractional calculus (FC) have been extensively used in the development of classical mechanics (Newtonian, and the latter sophisticated Lagrangian and Hamiltonian formulations), electromagnetics, nonrelativistic as well as relativistic quantum mechanics, quantum field theory, and statistical mechanics leading to their wide-ranging applications in atomic, molecular, nuclear, particle, and metal cluster physics. As a mathematical tool, FC also finds numerous applications in the description of anomalous diffusion in porous media, dense polymer solutions, composite heterogeneous films, etc; in understanding behaviour of viscoelastic materials, colloidal systems, magneto-rheological fluids, and amorphous semiconductors; in wave propagation in media with long-range interaction; in statistical mechanics of systems with long-range power-law interactions; in signal and image processing; in analysis and synthesis of different types of control systems; in the description of physics of biological structures and living organisms (e.g., DNA dynamics, protein folding, modeling for neuron activity, etc); in developing models for description of environment; and in mathematical modelling of economic processes and population growth with memory effects.

At the elementary physics level, many workers have used FC to discuss topics like Newton's second law of motion and equations of motion, a body falling under gravity in a viscous medium, projectile motion, simple oscillator, Fourier law of thermal conduction, Newton's law of cooling, Ohm's law and RC, RL circuits, and physics of decay / growth / relaxation processes. In this article, we describe harmonic oscillator in the framework of FC and bring out the memory effects in it.

## 2. Fractional Oscillator

A particle or an object executing repeated back and forth motion after being displaced from its equilibrium position in such a way that the restoring force is always directed towards the point of equilibrium and its magnitude is proportional to the displacement, is called a mechanical harmonic oscillator (HO). In one dimension, for a body of mass $m$ having instantaneous displacement $x(t)$ with equilibrium or centre point at $x = 0$, such a motion (known as simple harmonic motion) is described by the differential equation,

$$m\frac{d^2x(t)}{dt^2} = -kx(t). \tag{2}$$

Here, the negative sign takes care of the fact that the directions of force and the displacement are opposite to each other. The parameter k, the force per unit displacement, is called force constant. In case, the restoring force depends on higher powers of displacement, then the oscillator is said to be anharmonic. If the oscillating system experiences a friction-like force (which is generally always present in real systems) that causes continuous dissipation of energy, then it is referred to as a damped oscillator. Furthermore, an oscillator subjected to a time dependent external force is known as a forced or driven oscillator. The basic equation (2) is accordingly amended to accommodate relevant additional terms.

In general, any system, which need not necessarily be a material object and may be something like electric or magnetic field, that can be described by an expression analogous to that in Eq. (2) or its modified version, is said to be oscillatory. Besides everyday life examples of swings, cradles, vehicle shock absorbers, musical instruments, process of hearing, etc., considered in its general sense, an oscillator (classical or quantum one- or three- dimensional and their further modifications) is ubiquitous in physics and finds wide-range applications in developing theoretical models for different phenomena in almost all branches of physics. Some typical examples are simple, compound and torsional pendulums; spring-mass system; vibrating tuning forks, strings and gas columns; electric LC and LCR circuits; electronic oscillators; vibration of atoms in molecules; vibration of lattice atoms and molecules in solids leading to creation of phonons; modelling nuclear collective motion; basics of quantum field theory; and so on.

Replacing the ordinary second-order derivative by a Caputo FD of arbitrary order $\alpha$ ($1 < \alpha \leq 2$), Eq (2) can be written as

$$m_F D^\alpha x(t) = -kx(t), \tag{3}$$

This is fractional or generalized form of the equation of motion of a HO. To keep the meaning of $x(t)$ and k same as in Eq. (2), $m_F$ must have dimension $MT^{\alpha-2}$. Mass parameter $m_F$ becomes mass $m$ for index value $\alpha = 2$. We solve Eq. (3), using Laplace transform method, wherein a function $f(t)$ is converted into a function $F(s)$ of complex frequency $s$ through

$$F(s) \equiv \mathcal{L}\{f(t)\} = \int_0^\infty e^{-st} f(t) dt. \tag{4}$$

Taking LT of both sides of Eq. (3), using the fact that

$$\mathcal{L}\{D^\alpha f(t)\} = s^\alpha F(s) - s^{\alpha-1} f(0) - s^{\alpha-2} f^{(1)}(0), \tag{5}$$

with $f^{(1)}(0)$ as the first derivative of $f(t)$ at $t = 0$; we get

$$X(s) = \frac{x(0)s^{\alpha-1}}{\left(s^\alpha + \frac{k}{m_F}\right)} + \frac{v(0)s^{\alpha-2}}{\left(s^\alpha + \frac{k}{m_F}\right)}. \tag{6}$$

Inverse LT of Eq. (6) gives analytical expression for instantaneous displacement as

$$x(t) = x(0) E_{\alpha,1}\left(-\frac{k}{m_F} t^\alpha\right) + v(0) t E_{\alpha,2}\left(-\frac{k}{m_F} t^\alpha\right). \tag{7}$$

where $E_{\alpha,1}$ and $E_{\alpha,2}$ are two-parameter Mittag-Leffler functions, in general, defined by

$$E_{\rho_1,\rho_2}(z) = \sum_{n=0}^\infty \frac{z^n}{\Gamma(n\rho_1 + \rho_2)}, \qquad (\rho_1, \rho_2 > 0), \tag{8}$$

In particular,

$$E_{2,1}(-z^2) = \sum_{n=0}^{\infty} \frac{(-z^2)^n}{\Gamma(2n+1)} = \sum_{n=0}^{\infty} \frac{(-1)^n}{(2n)!} z^{2n} = \cos(z), \tag{9}$$

and

$$E_{2,2}(-z^2) = \sum_{n=0}^{\infty} \frac{(-z^2)^n}{\Gamma(2n+2)} = \frac{1}{z}\sum_{n=0}^{\infty} \frac{(-1)^n}{(2n+1)!} z^{2n+1} = \frac{\sin(z)}{z}. \tag{10}$$

It may be mentioned that

$$\mathcal{L}\{t^{\rho_1 l + \rho_2 - 1} E^{(l)}_{\rho_1,\rho_2}(\pm a t^{\rho_1})\} = \frac{l! \, s^{\rho_1 - \rho_2}}{(s^{\rho_1} \mp a)^{l+1}}, \qquad Re(s) > |a|^{1/\rho_1} \tag{11}$$

where $E^{(l)}_{\rho_1,\rho_2}(y) = \frac{d^l}{dy^l} E_{\rho_1,\rho_2}(y)$.

Coming to the expression for $x(t)$, we get only one of the two terms if either $v(0) = 0$ or $x(0) = 0$. Note that the condition $v(0) = 0, x(0) \neq 0$ implies that the oscillator is at the turning point at $t = 0$ and has amplitude $x(0)$. In contrast, the initial condition $x(0) = 0, v(0) \neq 0$ corresponds to the situation that the oscillator starts its motion from the equilibrium point at the origin with velocity $v(0)$. Assuming the first condition, we get

$$x(t) = x(0) E_{\alpha,1}\left(-\frac{k}{m_F} t^\alpha\right). \tag{12}$$

It is worth mentioning that in the framework of FC, linear momentum at any time $t$ is expressed in terms of position at that time by

$$p(t) = m_F \frac{d^{\alpha/2} x(t)}{dt^{\alpha/2}}, \qquad 1 < \alpha \leq 2; \tag{13}$$

Note that for $\alpha = 2$, this reduces to the conventional expression for instantaneous momentum. Using Eq. (12), we have

$$p(t) = m_F \frac{d^{\frac{\alpha}{2}}\left[x(0) E_{\alpha,1}\left(-\frac{k}{m_F} t^\alpha\right)\right]}{dt^{\frac{\alpha}{2}}} = m_F x(0) \sum_{n=0}^{\infty} \frac{\left(-\frac{k}{m_F}\right)^n}{\Gamma(n\beta+1)} D^{\frac{\alpha}{2}} t^{\alpha n}$$

$$= m_F x(0) \sum_{n=0}^{\infty} \left(-\frac{k}{m_F}\right)^n \frac{t^{\alpha(n-1/2)}}{\Gamma(n\alpha+1-\alpha/2)} = -m_F x(0) \frac{k}{m_F} t^{\alpha/2} E_{\alpha,1+\alpha/2}\left(-\frac{k}{m_F} t^\alpha\right). \tag{14}$$

In view of this, the total mechanical energy of the fractional oscillator at time $t$, viz. $\mathbb{E}(t) = (p^2(t)/2m_F) + (kx^2(t)/2)$, is given by

$$\mathbb{E}(t) = \frac{1}{2} m_F x^2(0) \left(\frac{k}{m_F}\right)^2 t^\alpha \left[E_{\alpha,1+\alpha/2}\left(-\frac{k}{m_F} t^\alpha\right)\right]^2 + \frac{1}{2} k x^2(0) \left[E_{\alpha,1}\left(-\frac{k}{m_F} t^\alpha\right)\right]^2. \tag{15}$$

Obviously, $\mathbb{E}(t)$ varies with time implying that the total mechanical energy of a fractional oscillator is not a constant or is not conserved.

Furthermore, for maximum allowed value of $\alpha$, i.e., $\alpha = 2$, using Eq. (9), we get from Eq. (12),

$$x(t) = x(0) \cos\left(\sqrt{\frac{k}{m}}\, t\right). \tag{16}$$

which gives displacement of a conventional HO with natural angular frequency $\omega = \sqrt{\frac{k}{m}}$. Next, substituting $\alpha = 2$ into Eqs. (14) and (15), we get using Eqs. (10) and (9),

$$p(t) = -m\, x(0) \frac{k}{m} t E_{2,2}\left(-\frac{k}{m_F} t^2\right) = -m\, x(0) \sqrt{\frac{k}{m}} \sin\left(\sqrt{\frac{k}{m}}\, t\right), \tag{17}$$

And

$$\mathbb{E}(t) = \frac{1}{2}mx^2(0)\left(\frac{k}{m}\right)^2 t^2 [E_{2,2}\left(-\frac{k}{m}t^2\right)]^2 + \frac{1}{2}kx^2(0)[E_{2,1}\left(-\frac{k}{m}t^2\right)]^2 = \frac{1}{2}kx^2(0). \quad (18)$$

Note that expression in Eq. (17) is the same as is obtained by taking $p(t) = m\frac{dx(t)}{dt}$. Also, Eq. (18) brings out conservation of mechanical energy for the conventional HO, which is in contrast with the finding in equation (15) for a fractional oscillator.

The time dependence of $x(t)$ and $\mathbb{E}(t)$ for an oscillator with $\sqrt{\frac{k}{m_F}} = 1.0$ rad $s^{-\alpha/2}$, as given by Eqs. (12) and (15), respectively, for different $\alpha$ values are projected in *Figures* 1 and 2. The plots of $\mathbb{E}(t)$ for $\alpha = 1.2$ and 1.5 are quite close to each other so that only thr former is shown in the figure. A look at *Figure* 1 reveals that for $\alpha = 2$, the plot represents a cosine variation so that displacement (of the standard oscillator) is periodic and has same amplitude. However, for $1 < \alpha < 2$, the system makes a finite number of oscillations with decreasing amplitudes and finally decays to $x(t) = 0$ position. This attenuation of oscillations is similar to that observed in a damped HO. Moreover, this effect becomes more prominent as $\alpha$ decreases; the number of zeros in the $x(t)$ plots is quite small for $\alpha$ close to 1 (implying quite rapidly damped oscillations). In the case of total mechanical energy (*Figure* 2), the graph for $\alpha = 2$ corresponds to constancy of $\mathbb{E}(t)$ and, hence, conservation of energy, while the $\alpha < 2$ curves exhibit decrease in energy with time as for the damped or dissipative motion. Once again, the decrease in the curve is faster for lower values of $\alpha$. Thus, a fractional oscillator ($1 < \alpha < 2$) behaves as a damped oscillator even though there is no resistive medium and this damping is absent for $\alpha = 2$, which corresponds to a simple HO. In other words, the past-history of the oscillator with $1 < \alpha < 2$ influences its motion by producing dampening effect as if it were interacting with itself – **memory effect**. Accordingly, it is usually referred to as fractional-order intrinsic damping. Since the effect becomes more pronounced with decrease in $\alpha$, the order of fractional derivative provides a measure of memory effects. With a view to look at this feature more minutely, we recall that damping and hence dissipation in a standard HO is obtained when a velocity-dependent term $(-\gamma\frac{dx(t)}{dt})$ is added to the restoring force term on the right side of Eq. (2) so that this is external in nature. It is indeed interesting to note that though we started with zeroth order (i.e., no) derivative in the restoring force in Eq. (3), we have also got an effect associated with first-order derivative of $x(t)$. This means that the fractional derivative on the left side of this equation has led to a behaviour which is a mixture of zeroth order and first order derivatives. The inclusion of (ubiquitous) dissipation effect in a natural way in the description makes it closer to reality, and, thus, brings out the importance of FC as a logical and comprehensive tool for modeling real systems by incorporating memory effects.

It is worth pointing out that Rekhviashvili et al [6] studied vibrations of a free piezoelectric plate under standard laboratory conditions and found that their experimental data was best accounted for with $\alpha = 1.998$ (reasonably small memory effect).

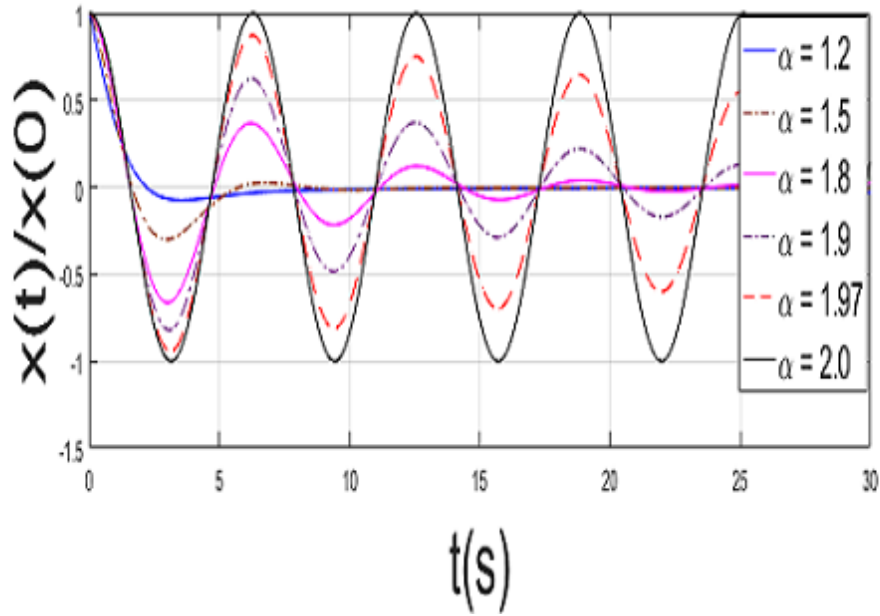

**Figure 1.** Graphical depiction of position of oscillator ($\sqrt{\frac{k}{m_F}} = 1.0$ rad $s^{-\alpha/2}$) as function of time for different values of $\alpha$.

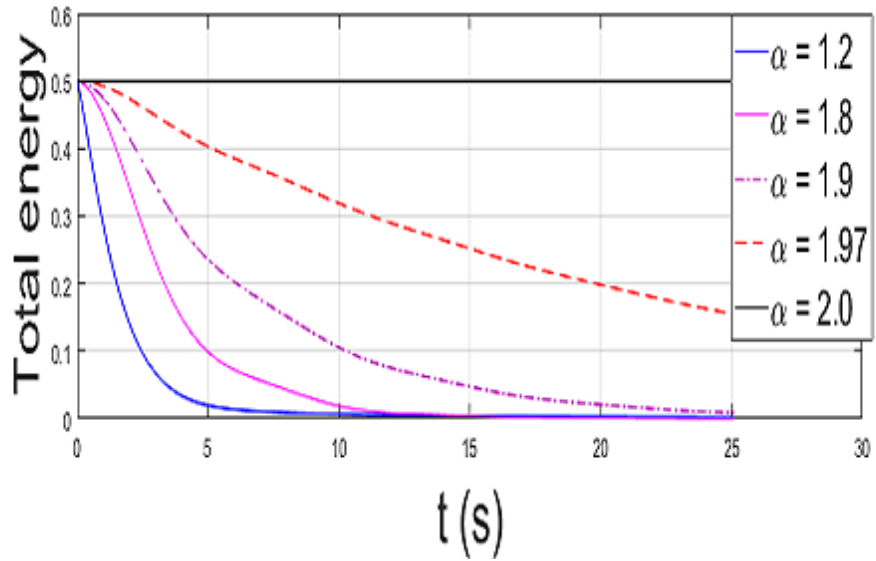

**Figure 2.** Dependence of total mechanical energy of oscillator ($\sqrt{\frac{k}{m_F}} = 1.0$ rad $s^{-\alpha/2}$) on time for different values of $\alpha$.

**Acknowledgement**

**Suggested Reading**